\documentclass[nohyper,11pt,letterpaper]{JHEP3}
\usepackage[dvips]{epsfig}
\usepackage{amsmath}

\newcommand{\be}{\begin{equation}}
\newcommand{\ee}{\end{equation}}
\newcommand{\ben}{\begin{eqnarray}}
\newcommand{\een}{\end{eqnarray}}
\newcommand{\ba}{\begin{eqnarray}}
\newcommand{\ea}{\end{eqnarray}}
\newcommand{\nn}{\nonumber \\}

\newcommand{\beq}{\begin{equation}}
\newcommand{\eeq}{\end{equation}}

\newcommand{\ie}{{\it i.e.}}
\newcommand{\etc}{{\it etc}}

\newcommand{\bN}{${\bf N}$ }
\newcommand{\sun}{${\bf SU(N)}$ }
\newcommand{\db}{${\bf D3}$ }

\newcommand{\ra}{\rightarrow}

\newcommand{\bi}{\begin{itemize}}
\newcommand{\ei}{\end{itemize}}
\newcommand{\ii}{\item}

\newcommand{\gb}{Gauss-Bonnet term}

\newcommand{\integrat}{\int_{\cal M} d^5x \sqrt{-g}}
\newcommand{\alp}{\alpha'}

\newcommand{\eom}{equation of motion}

\newcommand{\lb}{\left(}
\newcommand{\rb}{\right)}
\newcommand{\Ncg}{16 \pi G_5}

\newcommand{\pa}{\partial}
\newcommand{\mn}{\mu \nu}
\newcommand{\rl}{{r^2 \over L^2}}
\newcommand{\AL}{{\alpha' \over L^2}}
\newcommand{\B}{{\beta_i^{(0)}}}
\newcommand{\bo}{{b^{(0)}}}
\newcommand{\sig}{\sigma^{(0)}_{ij}}
\newcommand{\fac}{\lb 1- {\alpha' \over L^2} \rb}

\newcommand{\ltb}{\left [}
\newcommand{\rtb}{\right ]}
\newcommand{\bc}{\begin{center}}
\newcommand{\ec}{\end{center}}

\newcommand{\ga}{\gamma}

\title{Higher Derivative Corrections to Locally Black Brane Metrics}

\author {Suvankar Dutta\\
Harish-Chandra Research Institute, Chhatnag Road, Jhusi,
  Allahabad 211 019, INDIA\\
E-mail:\ \ {\bf suvankar@mri.ernet.in}}

\abstract{In this paper we generalize the construction of locally
  boosted black brane space time to higher derivative gravities.  We
  consider the Gauss-Bonnet term (with coefficient $\alp$) as a toy
  example. We find the solution to the $\alp$ corrected Einstein
  equations to first order in the boundary derivative expansion. This
  allows us to find the $\alp$ corrections to the boundary stress
  tensor in the presence of the Gauss-Bonnet term in the bulk
  action. We therefore obtain the ratio of shear viscosity to entropy
  which agrees with other methods of computation in the literature.}


\preprint{}

\begin{document}{\vskip 1cm}

\section{Introduction} \label{intro}

The conjectured duality between thermal gauge theory and gravity in
one higher dimensional AdS spacetime is a useful tool to extract
information about strongly coupled large $\bN$ gauge theory. For
example different thermodynamic variables like free energy, entropy
\etc \ of large $\bN$ $\sun$ gauge theory in ${\bf (3+1)}$ dimensions
in large 't Hooft coupling regime can be conjectured by calculating
the thermodynamic variables of black hole in five dimensional ${\bf
AdS}$ spacetime.

If we consider a black object with translation invariant horizon, for
example black ${\bf D3}$ brane geometry, one can also discuss
hydrodynamics - long wave length deviation (low frequency fluctuation)
from thermal equilibrium. In addition to the thermodynamic quantities
the black brane is also characterised by the hydrodynamic parameters
like viscosity, diffusion constant \etc. The black $\db$ brane
geometry with low energy fluctuations (\ie \ with hydrodynamic
behaviour) is dual to some finite temperature gauge theory plasma
living on boundary with hydrodynamic fluctuations.  Therefore studying
the hydrodynamic properties of strongly coupled gauge theory plasma
using the $AdS/CFT$ duality is an interesting subject of current
research. The energy momentum tensor of a relativistic viscous
conformal fluid is given by (up to first order in derivative
expansion),
\be \label{genemten}
T_{\mn} = (e + p) u_{\mu} u_{\nu}+ p \eta_{\mn} - 2 \eta \sigma_{\mn}
\ee
where $u_{\mu}$ is fluid 4-velocity with $u_{\mu}u^{\mu} =-1$, $e$ is
energy, $p$ is pressure and $\eta$ is shear viscosity
coefficient. $\sigma_{\mn}$ is defined in \ref{sigmamn}. Conformal
invariance implies that $e=3p$. The first attempt to study
hydrodynamics via AdS/CFT was \cite{pss}.  The authors related the
shear viscosity coefficient $\eta$ of strongly coupled ${\cal N}=4$
gauge theory plasma in large \bN limit with the absorption
cross-section of low energy gravitons by black \db brane. Other
hydrodynamic quantities like speed of sound, diffusion constants,
drag force on quarks \etc \ can also be computed in the context of
$AdS/CFT$ (See \cite{hydro} for review and the references therein).

If we consider the 't Hooft coupling to be very large but finite, then
we have to include the string theory contributions to thermodynamic
and hydrodynamic quantities, \ie \ we need to improve the supergravity
results by including the higher derivative terms in the action. The
higher derivative (string theory) corrections to shear viscosity have
been calculated in \cite{bls},\cite{gb1},\cite{gb2} \footnote{See also
\cite{buchelhydro}.}. There exists a ``viscosity bound conjecture''
\cite{kss} which states that the viscosity to entropy ratio ${\eta
\over s}$ has a lower bound,
\be 
{\eta \over s} \geq {1\over 4 \pi} 
\ee 
for all relativistic quantum field theories at finite temperature. 
In fact in presence of the ${\bf R^4}$ term
which is the first higher derivative correction appears in type II
string theory the viscosity entropy ratio is greater than ${1\over
4\pi}$. But the presence of Gauss-Bonnet term in the Lagrangian seems
to violate the ``viscosity bound conjecture''. Interested readers are
referred to follow \cite{gb1}, \cite{gb2}, \cite{gb3} for detailed
discussions. In all the cases the transport coefficients have been
determined either by the Kubo formula (graviton absorption) or by
quasi-normal mode calculation \cite{quasi1},\cite{quasi2} or using the
membrane paradigm approach \cite{horihydro}.

Recently \footnote{Also in \cite{secondhydro2}, authors have discussed
  second order hydrodynamics for conformal fluid. See also
  \cite{Natsuume}, \cite{rams} and \cite{gubser1} for related
  discussion.} in \cite{shiraj} the authors have developed an elegant
  systematic framework to construct the nonlinear fluid dynamics,
  order by order in boundary derivative expansion. The five
  dimensional Einstein equations with a negative cosmological constant
  with appropriate boundary conditions can be reduced to nonlinear
  equations of fluid dynamics. In this small note we have generalised
  the construction of local black brane geometry to higher derivative
  gravity. We started with Gauss-Bonnet term as a toy model.  We have
  found the solution of $\alp$ corrected Einstein equations only up to
  first order in boundary derivative expansion. Once we obtained the
  $\alp$ corrected geometry we calculate the boundary stress tensor up
  to first order in derivative expansion. From the expression of
  stress tensor one can read the $\alp$ correction to shear viscosity
  coefficient. We have also calculated the ratio of shear viscosity to
  entropy and the result agrees with existing results in literature
  calculated in other ways \cite{gb1}, \cite{gb2}.  Though we have
  found the correction to the metric up to first order in derivative
  expansion, but it would be interesting to find the corrections to
  the metric and stress tensor up to second order in derivative
  expansion \cite{nbsd1}.

We proceed in the same way of \cite{shiraj}. The presence of
Gauss-Bonnet term preserves all the symmetries of $AdS_5$
spacetime. We have also worked in the Eddington-Finkelstein
coordinate. Our solutions are also non-singular away from $r=0,$ and
specially at the location of the horizon.


The plan of the paper is following. In section \ref{rev} we will very
briefly sketch the calculation frame work of \cite{shiraj} and in
section \ref{calcu} we will display our results. We finish our paper
with some concluding remarks (section \ref{disco}).

\section{Fluid Dynamics From Gravity: The Computational Framework} \label{rev}

In this section we will briefly sketch the working procedure of
\cite{shiraj}. For detailed discussion readers are referred to the
original paper. We will also skip the technical details in this
section. 

\bi
\ii
Consider the Einstein-Hilbert action with negative cosmological
constant
\be
I = -{1 \over \Ncg} \int d^5x \sqrt{-g}\lb R + {12 \over L^2} \rb 
\ee
where $L$ is the radius of $AdS$ space.
\ii
The equation of motions are given by \footnote{$x^M = \{v, r,
\vec x\}$.}
\be
E_{MN} = R_{MN} -{1\over 2} R g_{MN} - {6\over L^2} g_{MN}=0.
\ee
\ii
There exists a class of solutions to these equations of motion given
by the ``boosted black branes'' \footnote{$x^{\mu} = \{v,\vec x
  \}$.},
\be
ds^2 = -2 u_{\mu}dx^{\mu}dr -\rl f(b r) u_{\mu} u_{\nu}
dx^{\mu}dx^{\nu} + \rl \lb u_{\mu} u_{\nu} + \eta_{\mn} \rb dx^{\mu}
dx^{\nu}
\ee
with, 
\ben
f(r)&=& 1- {1 \over r^4}, \nn
u_v&=& - \ga \nn
and \ \ u_i&=& \ga \beta_i
\een
where, $\ga = 1/\sqrt{1-\vec \beta^2}$.
\ii
Putting the values of $u_{\mu}$'s the metric can also be written as,
\ben \label{boostmetric}
ds^2 = && 2 \ga dv dr - {r^2\over L^2} \ga^2 f(br) dv^2 + \rl dx^i dx^i \nn
&+& \rl (\ga^2-1) dv^2 
- 2 \ga \beta_i dx^i dr - 2 \rl \ga^2 (1- f(br)) \beta_i dx^i dv \nn
&+&
\rl \ga^2 (1-
f(br)) \beta_i \beta_j dx^i dx^j.
\een 
The solution is parametrised by four constant parameters $b$ and
$\beta_i$'s. 
\ii
The black brane horizon is located at $r_H=1/b$ and the temperature of
this black brane is given by,
\be \label{temp0}
T= {1 \over \pi b L^2}.
\ee
\ii
Consider the metric \ref{boostmetric} and replace the constant
parameters $b$ and $\beta_i$'s by slowly varying functions $b(x^{\mu})$
and $\beta_i(x^{\mu})$'s of boundary coordinates $x^{\mu}$
\ben \label{boostmetric2}
ds^2 =&& 2\ga dv dr - {r^2\over L^2} \ga^2 f(b(x^{\alpha})r) dv^2
+ \rl dx^i dx^i \nn 
&+& \rl (\ga^2 -1)dv^2 - 2 \ga \beta_i(x^{\alpha}) dx^i dr - 2
\rl\ga^2 (1-
f(b(x^{\alpha})r)) \beta_i(x^{\alpha}) dx^i dv \nn
&+&\rl \ga^2 (1-
f(b(x^{\alpha})r)) \beta_i(x^{\alpha}) \beta_j(x^{\alpha}) dx^i dx^j.
\een  
We will call this metric $g^{(0)}(b(x^{\alpha}),\beta_i(x^{\alpha}))$.
\ii
In general the metric \ref{boostmetric2} is not a solution to Einstein
equations unless one adds some corrections to the metric and also the
parameters $b(x^{\alpha}),\beta_i(x^{\alpha})$ satisfy some set of
equations, which turn out to be the equations of boundary fluid
mechanics.
\ii
Write the parameters $b(x^{\alpha})$ and $\beta_i(x^{\alpha})$ and the
metric as a derivative expansion of the parameters. Up to first order
in derivative expansion,
\be \label{metexpn1st}
g= g^{(0)}(b(x^{\alpha}),\beta_i(x^{\alpha})) + \epsilon
g^{(1)}(b(x^{\alpha}),\beta_i(x^{\alpha})),
\ee
\be \label{bexpan}
b(x^{\alpha}) = b^{(0)}(x^{\alpha})
\ee
and
\be \label{betaexpn}
\beta_i(x^{\alpha}) = \beta^{(0)}_i(x^{\alpha})
\ee
where $\epsilon$ is a dimension less parameter whose power counts the
number of (boundary)spacetime derivatives acting on the
parameters. Since $b^{(1)}(x^{\alpha})$ and
$\beta^{(1)}_i(x^{\alpha})$ do not enter in to the first order
equation of motions, we have kept the expansion for $b$ and $\beta_i$'s
up to leading order.
\ii
In general one can write the metric and parameters as power series of
$\epsilon$. Then plug the metric in Einstein equations and solve the
metric and the parameters order by order (in $\epsilon$).  For example
in our case since we are interested up to first order, we will plug the
metric in Einstein equations and solve for $g^{(1)}$ and the constraint
equations imply some relations between the zero$^{th}$ order
parameters. We will work in a particular gauge,
\be
Tr ((g^{(0)})^{-1} g^{(1)})=0.
\ee
\ii
After finding the metric with first order fluctuations one can find
the boundary stress tensor (using the
definition given in \cite{bala-kraus}, \cite{myers}). The form of the
boundary stress up to first order in derivative expansion is given by,
\be
16 \pi G_5 T_{\mn} = {T^4
 \pi^4L^3 } \lb4 u_{\mu} u_{\nu} +
\eta_{\mn}\rb - {2 T^3 \pi^3 L^3} \sigma_{\mn},
\ee
where $\sigma_{\mn}$ is given by,
\be \label{sigmamn}
\sigma_{\mn} = P_{\mu}^{\alpha} P_{\nu}^{\beta} \pa_{( \alpha}
u_{\beta )} -{1\over 3} P_{\mn} \pa_{\alpha}u^{\alpha}
\ee
and $P_{\mn} = u_{\mu}u_{\nu}+ \eta_{\mn}$.
\ei

\section {Higher Derivative Correction to First Order Hydrodynamics}
\label{calcu}

In this section we will explicitly show how one can generalise this
procedure to higher derivative action. We will consider the
Gauss-Bonnet action as a toy example. We will show how the first order
metric and constraint relations receive $\alp$ corrections. As a
result the boundary stress tensor also receives an $\alp$
correction. We have considered the Gauss-Bonnet term as a perturbation
and hence our metric and stress tensor is correct up to first order in
$\alp$.

\subsection {The Action} \label{action}

We will start with following action,
\ben \label{gbaction} 
I = &-& {1\over 16 \pi G_5}\integrat \left(R+{12
\over b^2}\right)-{\alp \over 16 \pi G_5}\integrat L_{GB}
\een 
where, 
\be \label{gbterm} 
L_{GB}=R_{MNPQ}R^{MNPQ} - 4 R_{MN} R^{MN} + R^2.
\ee 
The
equation of motion is
\ben \label{gbeom}
E_{MN}=R_{MN} &-& {1 \over2}R g_{MN} -{6 \over L^2}g_{MN} - {\alp
\over 2}g_{MN} L_{G.B.}\nn &+& {2 \alp } \left(R_{M P Q L} R_{N}^{\ PQL}
- 2 R^{PQ} R_{M P N Q} -2 R_{M}^{\ Q} R_{NQ} + R R_{MN} \right)=0.
\een

\subsection{The Counterterm and Boundary CFT Stress Tensor}
\label{sec:stressten}

As usual in gravity theories, the action (\ref{gbaction}) should be
supplemented with suitable boundary terms, for a well-defined
variational principle.  For Einstein gravity, one consider the
Gibbons-Hawking surface term \cite{gh}
\begin{equation}\label{IGH}
I_{GH}^{E}=-{1\over 8\pi G}\int_{\partial
\mathcal{M}}d^{4}x\sqrt{-\gamma }K~,
\end{equation}
where $\gamma _{\mu \nu }$ and $K$ are the induced metric and the
trace of the extrinsic curvature of the boundary, respectively.
\be
\gamma_{\mu \nu}= g_{\mu \nu}-n_{\mu}n_{\nu},
\ee
\be \label{K} 
K_{\mu \nu} = - {1 \over 2} \lb
\nabla_{\mu} n_{\nu} + \nabla_{\nu} n_{\mu}\rb 
\ee 
and $n_{\mu}$ is unit outward normal vector to the asymptotic
boundary hypersurface.

A similar term occurs for Gauss-Bonnet gravity
and reads \cite{Myers:1987yn,Davis:2002gn} 
\begin{equation}
I_{b}^{(GB)}=-\frac{1}{8\pi G}\int_{\partial \mathcal{M}}d^{4}x\sqrt{-\gamma }%
\left\{ 2\alpha'\left( J-2E_{\mn}^{(1)}K^{\mn}\right) \right\}~,
\label{IGB}
\end{equation}
with $ E_{\mn}^{(1)}$ is the 
four-dimensional Einstein tensor 
of the metric $\gamma _{\mn}$ and $J$ is the
trace of
\be \label{J}
J_{\mu \nu}={1\over 3} \lb 2 K K_{\mu \rho}K^{\rho}_{\nu} + K_{\rho
\sigma} K^{\rho \sigma} K_{\mu \nu} - 2 K_{\mu \rho} K^{\rho \sigma}
K_{\sigma \nu} - K^2 K_{\mu \nu}\rb . 
\ee
Variation of the action $I+I_{b}^{(E)}+I_{b}^{(GB)}$
now gives an expression
which does not contain
normal derivatives of $\delta g_{\mn}$.

It is well known that the total action has diverges even at tree
level. The divergence arise from integrating over the infinite volume
of spacetime.  We regularise this divergence by using the procedure
proposed by Balasubramanian and Kraus \cite{bala-kraus}, which
furnishes a method for calculating gravitational action and conserved
quantities without reliance on any reference spacetime.  This
technique was inspired by AdS/CFT correspondence and consists of
adding suitable counter terms $I_{ct}$ to the action of the theory in
order to ensure the finiteness of the boundary stress tensor
\cite{Brown}.

We have found that the on-shell action
can be regularised 
by the following counter term (see also \cite{cardoso}).
\begin{eqnarray}\label{Ict1}
I_{\mathrm{ct}} &=& {1 \over 8 \pi G_5} \int{\partial
\mathcal{M}}d^{4}x\sqrt{-\gamma } \lb c_1 + {c_2 \over 2}
\mathcal {R} \rb 
\een
where $c_1$ and $c_2$ are functions of $\alp$. $\mathcal{R}$ is the
Ricci scalar made out if boundary metric $\gamma$. For flat boundary
geometry ($R^1 \times R^3$), $\mathcal{R}=0$ and the counterterm terns
out to be (up to first order in $\alp$),
\ben \label{Ict}
I_{\mathrm{ct}}&=&\frac{1}{8\pi G_5}\int_{\partial \mathcal{M}}
d^{4}x\sqrt{-\gamma } \lb{3 \over L^2} - {\alp \over L^3} \rb.
\end{eqnarray}

Varying the total action (which contains the  boundary terms 
(\ref{IGH}),(\ref{IGB}) and (\ref{Ict}))
with respect to the
boundary metric $\gamma_{\mn}$, we compute the 
divergence-free boundary stress-tensor
\begin{eqnarray}
S_{\mn}=\frac{1}{8 \pi G_5} \left( K_{\mn}-K\gamma_{\mn} + 2 {\alpha'}
(3J_{\mn}-J\gamma_{\mn}) -{3 \over L^2}\gamma_{\mn} + {\alp \over
L^3}\gamma_{\mn} \right).
\end{eqnarray} 

Since the CFT metric is given by,
\be \label{cftmetric}
h_{\mu \nu} = \lim_{\tilde R \rightarrow \infty}  {L^2 \over \tilde
  R^2} \gamma_{\mu \nu},
\ee
the boundary CFT stress tensor is given by \cite{myers},
\ben \label{stressten}
T_{\mu \nu} &=& \lim_{\tilde R \rightarrow \infty} {{\tilde R}^2\over
  L^2} S_{\mn} \nn
&=& \lim_{\tilde R \rightarrow \infty} {{\tilde R}^2\over L^2} {1
\over 8 \pi G_5}
\lb K_{\mu \nu} - K \gamma_{\mu \nu} + 2 \alp (3 J_{\mu \nu} - J
\gamma_{\mu \nu}) - {3 \over L} \gamma_{\mu \nu} + {\alp \over L^3}
\gamma_{\mu \nu} \rb 
\een
where $\tilde R $ is the cutoff in radial direction. The gauge theory
lives on the boundary
of the AdS space which is at $r= \tilde R$.


\subsection {Solution to the Equation of Motions} \label{solution}

The solution to the \eom \ is given by the metric \cite{sd},
\ben \label{metric}
ds^2 = 2 dv dr - {r^2 \over L^2} f(br) dv^2 + {r^2 \over L^2} d{x^i }d{x^i },
\een
where the function $f(br)$ is given by,
\be
f(br) = 1 - {1 \over (br)^4} + {2 \alp \over L^2} \lb1 + {1 \over
(br)^8}\rb.
\ee
The horizon radius $r_H$ is given by,
\be
r_H= {1-\alp \over b} 
\ee
and 
the temperature of this black brane is,
\be \label{temp} 
T= {f'(r_H) \over 4 \pi} = {1 \over \pi b L^2} \lb
1-{\alp \over L^2} \rb .  
\ee

By giving a coordinate transformation one can obtain the boosted black
brane metric which is of the following form $(u_{\mu}= (-\ga,\ga
\vec{\beta}))$,
\ben \label{boostmetricalp}
ds^2 = && 2 \ga dv dr - {r^2\over L^2} \ga^2 f(br) dv^2 + \rl dx^i dx^i \nn
&+& \rl (\ga^2-1) dv^2 
- 2 \ga \beta_i dx^i dr - 2 \rl \ga^2 (1- f(br)) \beta_i dx^i dv \nn
&+&
\rl \ga^2 (1-
f(br)) \beta_i \beta_j dx^i dx^j.
\een

The CFT metric is given by \ref{cftmetric}.
In $\alp \ra 0$ limit the CFT metric $h_{\mu \nu}$ is simply
the Minkowski metric $\eta_{\mu \nu}$. But when we include the $\alp$
correction then the CFT metric is no longer $\eta_{\mn}$. It
has the following form (for the metric \ref{metric}),
\be
h_{\mn}\ = \
\lb 
\begin{array}{cccc}
-1-{2\alp \over L^2} & \ \ 0\ \ & \ \ 0 \ \ & \ \ 0\\
0\ \  &\ \  1\ \  &\ \  0\ \  &\ \  0\\
0\ \ & \ \ 0\ \ &\ \ 1\ \ &\ \ 0\\
 0\ \  &\ \  0\ \  &\ \  0\ \  &\ \  1
\end{array} \rb. 
\ee
For metric \ref{boostmetricalp}, $h_{\mn}$ also has off-diagonal terms
proportional to $\alp$. 
If we want to keep the CFT metric to be $\eta_{\mn}$ then we
have to rescale either the time coordinate $v \ra (1 - \alp/L^2) \ v$ 
or the space coordinates $x^i \ra (1 + \alp/L^2) \ x^i$. Here we will
rescale the time coordinate 
\be
v=\lb 1 - {\alp \over L^2} \rb V.
\ee
In this rescaled coordinate $(V,\vec x)$ the metric \ref{metric} takes
the following form,
\be 
ds^2 = 2\lb 1 - \AL \rb dV dr  - \rl \tilde f(br) dV^2 + \rl dx^i
dx^i.
\ee
And the temperature has also been rescaled to,
\be
T= {1 \over \pi b L^2} \lb 1-{2 \alp \over L^2} \rb .
\ee
Now we can give a coordinate transformation (boost) and the boosted
black brane metric in ($V$,$\vec x$) coordinate becomes,
\ben \label{rescaledmet}
ds^2 = &&2 \ga \lb 1 - \AL \rb dV dr - \rl \ga^2 \tilde f(br) dV^2 + \rl dx^i
dx^i\nn 
&+& \rl (\ga^2-1) dV^2- 2  \ga \lb 1 - \AL \rb \beta_i dx^i dr 
- 2 \rl \ga^2 (1 - \tilde f(br))
\beta_i dx^i dV \nn
&+& \ga^2 \rl (1 - \tilde f(br)) \beta_i \beta_j dx^i dx^j,
\een
where 
\be
\tilde f(br) = \lb 1 - {2 \alp \over L^2} \rb f(br).
\ee
          
\subsection{First Order Metric} \label{1stordmet}

In this section we will determine the metric to the first order in the
derivative expansion. The metric and the parameters are given by Eq.
\ref{metexpn1st}, \ref{bexpan} and \ref{betaexpn}.

We will choose a coordinate to set $u_{\mu}=\{-1,0,0,0\}$ (\ie \ $u_V =
-1$, $\beta_i^{(0)} =0$,$\ga=1$) and $b^{(0)} =1$ at a given point
$x^{\mu}=0$\footnote{From now, our $x^{\mu} = \{V,\vec x\}$. }. Around
this point the velocities and the temperature fields can be expanded
up to first order in derivative,

\be \label{bexpn2}
b= 1 + x^{\mu}\pa_{\mu}\bo 
\ee
and
\be
\B= x^{\mu}\pa_{\mu}\B.
\ee
So the metric up to first order in derivatives is given by,
\ben \label{met1st}
ds^2= &2& \lb 1 - \AL \rb dV dr - \rl \tilde f(r) dV^2 + \rl dx^i
dx^i\nn
&-& 2 \lb 1 - \AL \rb x^{\mu} \pa_{\mu} \B dx^i dr \nn
&-& 2 \rl (1 - \tilde
 f(r))x^{\mu} \pa_{\mu} \B dx^i dV -{r^3 \over L^2} x^{\mu} \pa_{\mu}
 \bo \tilde f'(r) dV^2.
\een
This is the metric $g^{(0)}$ expanded up to first order in derivative.

As we explained earlier $g^{(0)}$ is not a solution to \eom. We have
to find the fluctuation metric $g^{(1)}$, which added to $g^{(0)}$
solves the \eom \ up to first order in derivative expansion. 

Because of spatial ${\bf SO(3)}$ symmetry of background black brane
metric, we can separately solve for the ${\bf SO(3)}$ scalars, ${\bf
SO(3)}$ vectors and ${\bf SO(3)}$ symmetric traceless components of
$g^{(1)}$.

\subsubsection{The Scalar Sector} \label{scalarsector}

The scalar components of $g^{(1)}$ are parametrised in the following
way,
\ben \label{scalarpara}
g^{(1)}_{ii}(r) &=& 3 \rl h_1(r) \ \ (sum \ over \ i)\nn 
g^{(1)}_{VV}(r) &=& {L^2 \over r^2} k_1(r)\nn
g^{(1)}_{Vr}(r) &=& - {3 \over 2} \lb 1 - \AL \rb h_1(r).
\een
Here we are working in the following gauge,
\be
Tr ((g^{(0)})^{-1} g^{(1)})=0.
\ee
The scalar Einstein equations (equations invariant under ${\bf SO(3)}$
rotations) are divided up into constraint and dynamical
equations. 
\bc
\underline{{\bf Constraint Equation 1}}
\ec
The first scalar constraint is 
\be
\rl \tilde f(r) E_{Vr} + E_{VV} = 0,
\ee
which evaluates to 
\ben \label{scalarconst1}
&&\lb 1 - {5 \alp \over L^2} \rb (\pa_V \bo - {1\over 3} \pa_i \B)=0\nn
&& i.e. \ \ \ \pa_V \bo - {1\over 3} \pa_i \B=0 .
\een
The constraint relation remains unchanged in presence of Gauss-Bonnet
correction. This relation is a consequence of the conservation of
boundary energy momentum tensor.
\bc
\underline{{\bf Constraint Equation 2}}
\ec

The second constraint equation is 
\be 
\rl \tilde f(r) E_{rr} + E_{Vr} = 0,
\ee
leads to
\be \label{scalarconst2}
12 r^3 h_1(r) + (3 r^4 -1) h_1'(r) - L^4 k_1'(r)= -2 L^2 r^2 \lb 1-
\AL \rb\pa_i \B . 
\ee
\bc
\underline{{\bf Dynamical Scalar Equation}}
\ec
In addition to these constraint equations we have to add one dynamical
equation. Like \cite{shiraj} we will add the following simplest
equation, 
\be \label{scalardyna}
E_{rr}=5 h_1'(r) + r h_1''(r) =0.
\ee

We will solve the equations \ref{scalarconst2} and \ref{scalardyna} to
find out the function $h_1(r)$ and $k_1(r)$. These two equations
satisfy all the regularity and normalisation conditions explained in
\cite{shiraj}. Hence the solutions are given by,
\ben \label{scalarsol}
h_1(r) &=& 0, \nn
k_1(r)&=& {2 \over 3} \rl \lb 1 - \AL \rb \pa_i \B.
\een
The constants appear in $h_1(r)$ and $k_1(r)$ can be set to zero
following the same argument in
\cite{shiraj}. So the scalar part of the fluctuation metric is given
by \footnote{$g_S^1, g_V^1, g_T^1$ are the scalar, vector and tensor
  part of the fluctuation metric $g^1$ respectively.},
\be
\lb g_S^{(1)} \rb_{\alpha \beta}dx^{\alpha}dx^{\beta} 
= {2\over 3} r \lb 1 - \AL \rb \pa_i
\B \ dV^2.
\ee

\subsubsection{The Vector Sector} \label{vecsector}

We will parametrise the vector part of the fluctuation metric as,
\be
\lb g_V^{(1)} \rb_{\alpha \beta} = 2 \rl (1 - \tilde f(r)) j^{(1)}_i
dV dx^i.
\ee
\bc
\underline{{\bf Constraint Equation 3}}
\ec
In this sector the constraint equation is given by,
\be
\rl \tilde f(r) E_ {ri} + E_{Vi} =0
\ee
which gives,
\ben \label{vecconst}
&& \lb 1 - {5 \alp \over L^2} \rb (\pa_i \bo - \pa_V \B)=0 \nn
&& i.e. \ \ \pa_i \bo - \pa_V \B=0.
\een
Again the presence of Gauss-Bonnet term does not have any effect on
this constraint equation which can be interpreted as a consequence of
conservation of boundary stress tensor.
\bc
\underline{{\bf Dynamical Equation for $j^{(1)}_i(r)$}}
\ec

The dynamical equation in vector sector is,
$E_{ri}=0.$ The equation for $j_i^{(1)}(r)$ turns out to be \footnote{when
  $j_i^{(1)}$'s or their derivatives appear with $\alp$ then we have set the
leading order (in $\alp$) values of $j_i^{(1)}$'s or their derivatives.},
\be
{d \over dr} \lb {1 \over r^3} {d \over dr} j^{(1)}_i(r) \rb = - 3
{L^2 \over r^2} \pa_V \B + \AL \lb {10 L^2 \over r^6} - {3 L^2 \over
r^2} \rb \pa_V \B. 
\ee
The solution is given by,
\be
j^{(1)}_i(r) = L^2 r^3 \ltb 1 + \AL \lb 1 + {2 \over r^4} \rb \rtb
\pa_V \B .
\ee 
The integration constants are set to zero since the solution should be
normalisable at the boundary and the stress tensor must be
renormalisable \cite{shiraj}. 
So the vector part of the fluctuation metric is given
by,
\be
\lb g_V^{(1)} \rb_{\alpha \beta}dx^{\alpha}dx^{\beta} 
= 2 r \pa_V \B \lb 1 - \AL \rb dV dx^i.
\ee


\subsubsection{The Tensor Sector} \label{tensector}

The ${\bf SO(3)}$ tensor part of $g^{(1)}$ can be parametrise in the
following way  
\be
\lb g_T^{(1)}\rb_{\alpha \beta} = \rl \alpha^{(1)}_{ij}dx^i dx^j
\ee
where $\alpha^{(1)}_{ij}$ is symmetric traceless $3 \times 3$
matrix.
\bc
\underline{{\bf Dynamical Equation for $\alpha^{(1)}_{ij}(r)$}}
\ec
Equation for $\alpha^{(1)}_{ij}(r)$ follows from the $E_{ij}=0.$ 
\ben \label{alphaeq}
{d \over dr} \lb (r^5 -r) {d \over dr} \alpha_{ij}^{(1)}(r) \rb = &-& 6
L^2 r^2 \sigma^{(0)}_{ij} \nn
&+& \AL {1 \over r^4} \lb 2 (9 + 5 r^8)
\alpha^{'(1)}_{ij}(r) \rb \nn 
&+& \AL {1 \over r^4} \lb  2 L^2 r^2 (-4 + 9 r^4) \sig + 2r (-3 + r^8)
\alpha^{''(1)}_{ij}(r) \rb 
\een
where,
\be
\sig= \pa_{(i}\beta^{(0)}_{j)} - {1\over 3} \delta_{ij} \pa_m
\beta_m^{(0)}.
\ee

The solution for $\alpha^{(1)}_{ij}(r)$ is given by \footnote{In the
  right hand side of Eq. \ref{alphaeq} we set the leading order
(in $\alp$) values of $\alpha_{ij}^{(1)}$ and their derivatives.},
\be
\lb g_T^{(1)}\rb_{\alpha \beta}dx^{\alpha} dx^{\beta} = 2  \lb
  r \lb 1 - \AL \rb - {1 \over 4 r^2 } \lb 1 - {8 \alp \over
 L^2} \rb \rb \sig dx^i dx^j.
\ee

\bc
\underline{{\bf Summary of $\alp$ corrected first order calculation}}
\ec

The $\alp$ corrected metric $g^{(0)} +g^{(1)}$ expanded up to first
order in boundary derivatives around some point $x^{\mu}$ where
$b^{(0)}=1$ and $u_{\mu} = \{-1, 0,0,0\}$ is given by,
\ben \label{met1stfinal}
ds^2= &2& \lb 1 - \AL \rb dV dr - \rl \tilde f(r) dV^2 + \rl dx^i
dx^i\nn
&-& 2 \lb 1 - \AL \rb x^{\mu} \pa_{\mu} \B dx^i dr \nn
&-& 2 \rl (1 - \tilde
 f(r))x^{\mu} \pa_{\mu} \B dx^i dV -{r^3 \over L^2} x^{\mu} \pa_{\mu}
 \bo \tilde f'(r) dV^2 \nn
&+& {2\over 3} r \lb 1 - \AL \rb \pa_i
\B \ dV^2 + 2 r \pa_V \B \lb 1 - \AL \rb dV dx^i \nn
&+& 2  \lb
  r \lb 1 - \AL \rb - {1 \over 4 r^2 } \lb 1 - {8 \alp \over
 L^2} \rb \rb \sig dx^i dx^j.
\een

\bc
\underline{{\bf Global solution to first order in derivative}}
\ec

The metric \ref{met1stfinal} has been calculated about $x^{\mu} =0$
assuming $\bo=1$ and $\B =0$. But one can also write the metric about
any point. The $\alp$ corrected global metric is given by,
\ben \label{globmet}
ds^2 =&& -\fac u_{\mu} dx^{\mu} dr - \rl \tilde f(b r) u_{\mu}
u_{\nu}dx^{\mu} dx^{\nu} + \rl P_{\mn}dx^{\mu} dx^{\nu} \nn 
&+& 2 r^2 b F(br) \sigma_{\mn} dx^{\mu} dx^{\nu} + {2\over 3} \fac r
u_{\mu} u_{\nu} \pa_{\lambda} u^{\lambda} dx^{\mu} dx^{\nu} \nn
&-& r \fac
u^{\lambda} \pa_{\lambda} (u_{\mu} u_{\nu}) dx^{\mu} dx^{\nu}
\een
where $F(r)$ is given by,
\be
F(r)= \lb
  {1 \over r} \lb 1 - \AL \rb - {1 \over 4 r^4 } \lb 1 - {8 \alp \over
 L^2} \rb \rb.
\ee


\subsection{The Energy Momentum Tensor} \label{emten}

Once we obtained the $\alp$ corrected metric up to first order
fluctuations, we can find the boundary stress tensor using the
definition \ref{stressten}. 
Different components are given by,
\ben 
16 \pi G_5 T_{VV} &=& {3 \over L^5 b^4}\lb 1 - {5 \alp \over L^2} \rb
= {3 T^4 \pi^4
L^3} \lb 1 + {3 \alp \over L^2} \rb \nn
16 \pi G_5 T_{ij} &=& {1 \over L^5 b^4}\lb 1 - {5 \alp \over L^2} \rb
\delta_{ij} 
- {2 \over L^3
b^3} \lb 1 - {11 \alp \over L^2} \rb \sig\nn
&=& { T^4
 \pi^4 L^3} \lb 1 + {3 \alp \over L^2} \rb \delta_{ij} - {2 T^3   
\pi^3 L^3} \lb 1  - {5 \alp \over L^2} \rb \sig\nn
16 \pi G_5 T_{Vi} &=& - {4 \over L^5}  \lb 1 - {5 \alp \over
L^2} \rb x^{\mu} \pa_{\mu} \B \nn
&=& - 4 { T^4   \pi^4 L^3}  \lb 1 + {3 \alp \over L^2} \rb x^{\mu}
\pa_{\mu} \B. 
\een

Here we have written the energy momentum tensor about for the metric
\ref{met1stfinal} \ie \ about $x^{\mu}=0$ assuming that $\bo =1$ and $\B
=0$ at the origin. But for the global metric \ref{globmet} the energy
momentum tensor can be written in a covariant way. Up to first order
in derivative expansion the energy momentum tensor has the form given
by Eq. \ref{genemten} (see \cite{loga}). Therefore in presence of
the \gb \ one can write the energy momentum tensor in a covariant form
in the following way,
\ben
16 \pi G_5 T_{\mn} &=& {1 \over L^5 b^4}\lb 1 - {5 \alp \over L^2} 
\rb \lb 4 u_{\mu} u_{\nu} +
\eta_{\mn}\rb - {2 \over L^3
b^3} \lb 1 - {11 \alp \over L^2} \rb \sigma_{\mn} \nn
&=& {T^4
 \pi^4L^3 } \lb 1 + {3 \alp \over L^2} \rb \lb4 u_{\mu} u_{\nu} +
\eta_{\mn}\rb - {2 T^3 \pi^3 L^3} \lb 1 - {5 \alp \over L^2}\rb \sigma_{\mn}.
\een
The shear viscosity coefficient is given by,
\be 
\eta = {1 \over 16 \pi G_5 L^3 b^3}\lb 1 - {11 \alp \over L^2} \rb = {
T^3 \pi^3 L^3 \over 16 \pi G_5} \lb 1 - {5 \alp \over L^2}\rb.
\ee
Entropy density is given by,
\ben
s= {S \over V_3} &=& {Area \over 4 G_5 V_3}\nn 
&=& {r_H^3 \over 4 L^3 G_5}\nn
&=&{1\over 4 L^3 b^3 G_5}\lb 1 - {3 \alp \over
L^2}\rb \nn
&=& {T^3 \pi^3 L^3 \over 4 G_5}\lb 1 +
{3 \alp \over L^2} \rb.
\een
where $V_3 = \int d^3x$.
Hence,
\be
{\eta \over s} = {1 \over 4 \pi} \lb 1 - {8 \alp \over L^2} \rb
\ee
which is in agreement with \cite{gb1}, \cite{gb2}.



\section{Discussion} \label{disco}

In this paper we have constructed the local black ${\bf D3}$ brane
geometry in presence of Gauss-Bonnet term in the bulk action up to
first order in derivative expansion. The local solution we found is
non-singular except at $r=0$.  We used the counter term method of
Balasubramanian and Kraus to find the boundary stress tensor , which
we have expanded up to first order in derivatives, for the $\alp$
corrected metric. From the expression of stress tensor one can read
the shear viscosity coefficient. We have found the $\alp$ correction
to the shear viscosity coefficient and also to the viscosity entropy
ratio. These results are in agreement with the existing results in the
literature.

Recently in \cite{shiraj2}, authors have demonstrated that given a
black brane geometry with regular event horizon the location of the
horizon in radial direction turns out to be a local function of fluid
dynamical variables evaluated at the corresponding points on the
boundary. In presence of this regular local event horizon they
constructed an appropriate area form on spatial section and then
taking the pull-back of this area form to the boundary they defined a
local entropy current for the dual field theory. The entropy current
has the following form
\be
4 G_5 L^3 b^3 J_S^{\mu} = u^{\mu} + {\cal O}(\epsilon^2).
\ee
The first order $({\cal O}(\epsilon))$ correction to entropy current
is zero. As mentioned in \cite{shiraj2} finding out the $\alp$
correction to entropy current would be an interesting problem to
solve. Since the relations Eq. \ref{scalarconst1} and
Eq. \ref{vecconst}, which follow from the conservation of energy
momentum tensor, remain unchanged (up to an overall factor),
therefore, given the $\alp$ corrected bulk metric \ref{globmet} it is
easy to check that the entropy current, up to ${\cal O}(\epsilon)$,
has the following form ,
\be
4 G_5 L^3 b^3 J_S^{\mu} = \lb 1 - {3\alp \over L^2} \rb u^{\mu} .
\ee
Again the entropy current does not receive any ${\cal O}(\epsilon)$
correction. So to find out the $\alp$ correction up to order ${\cal
O}(\epsilon^2)$ to the entropy current, one has to first find the
corrected metric up to ${\cal O}(\epsilon^2)$ \cite{nbsd1}.

Although we have considered only the four derivative terms in the
Lagrangian but it would be very interesting to generalise this idea
for any higher derivative gravity. From the string theory point of
view it would be nice to construct the corrected local black ${\bf
D3}$ brane geometry for ${\bf R^4}$ term \cite{nbsd2}.

\vspace{1cm}

\noindent
{\bf \large{Acknowledgement}}\\

We are thankful to R. Gopakumar for numerous helpful discussions. We
also thank N. Banerjee, R. Gupta and S. Minwalla for useful
discussions.

\end{document}